\documentstyle[epsfig,graphicx,preprint,aps]{revtex}
\tightenlines
\begin{document}
\draft
\title{Constraints on vector meson photoproduction
spin observables} 
\author{W.M. Kloet}
\address{ Department of Physics \& Astronomy, Rutgers University,
 Piscataway, New Jersey  08855-0849}
\author{Frank Tabakin }
\address{ Department of Physics \& Astronomy, University of Pittsburgh,
 Pittsburgh, Pennsylvania  15260}
\date{\today}
\maketitle
\ 
\begin{abstract}
Extraction of spin observables from vector meson
photoproduction on a nucleon target is described.  
Starting from density matrix elements in the vector meson's rest frame,  
we transform to spin
observables in the photon-nucleon c.m. frame.  
 Several constraints on the transformed density matrix and on
the spin observables follow from requiring that the
angular distribution and the density
matrix be positive definite.   A set of
constraints that are required in order to extract meaningful spin
observables from forthcoming data are enunciated.
\end{abstract}
\widetext

\section{INTRODUCTION}
\label{sec:introduction}
  In an earlier paper~\cite{KCT}, we emphasized that 
the angular distribution of the pseudoscalar  
 mesons that arise from the decay of photoproduced vector mesons 
does not depend on the vector meson's vector polarization, but only on 
its tensor polarization and that standard single and double spin 
observables need to be defined in the overall photon-nucleon 
center of mass frame.
It was also found that a simple description for the 
decay angular distribution in the $\gamma N$ c.m. frame
is obtained by using 
  the angle between the decay meson's velocity difference vector 
 and the direction of the photoproduced vector meson. 
  The main purpose of this paper is to formulate a procedure for 
extracting meaningful spin observables from the analysis of 
forthcoming vector meson photoproduction data to allow one 
to examine conventional spin observables.
 These spin observables are subject to known rules concerning 
their forward and backward angular behavior~\cite{PST}.
 The nodal structure of spin observables, 
 e.~g., their production angle dependence, might reveal important underlying
 dynamics such as baryonic and mesonic resonances.

Here we show how to extract meaningful spin observables
 under the assumption that analysis of the photoproduction of
 vector mesons will yield a vector meson rest-frame density matrix. 
   Since new data are not yet available,  we invoked older
   1968 Aachen et al. information~\cite{Aach} and found that
 some of their vector meson rest-frame density matrix results,
 when transformed 
 to spin observables in the photon-nucleon center of mass frame,
violated basic constraints and therefore need to be rejected.
   The grounds for that rejection was that some of their elements, 
 even including their stated uncertainties, yielded  non-positive, and
  therefore unacceptable angular distribution functions.   
  That observation, which we subsequently found to be related to a set of
constraints deduced by others earlier~\cite{Daboul,Minnaert}, led us to examine the various constraints based on the positivity of the density matrix.

In Sec. II,  we analyze the limits on the tensor polarization
 provided by the simple requirement that the angular distribution of
 the decay mesons be a positive definite function.
  Simple limits on the tensor polarizations follow from evaluating
 the decay angular distribution at selected angles. 
 In Sec. III,  the limits on observables due to
positivity of the density matrix are discussed.
  In Sec. IV,  constraints on the vector meson's density matrix due to
 Daboul's analysis~\cite{Daboul}
   using Schwarz inequalities,  are invoked and analyzed.
 The Schwarz inequalities described by Daboul 
 can be re-expressed as four separate conditions on spin observables.
  However, all but two of these conditions are already contained in the
 simple requirement that the angular distribution should be positive definite.
 The two remaining conditions involve not only
 the tensor polarization, but also 
the vector meson's vector polarization. These conditions 
could also be used to limit the vector meson's
 vector polarization.
Finally, in Sec.~V the method for extracting spin observables
 from actual data is outlined.

  Use of these  basic constraints should  
 be included in the fitting procedure to assure that general
 requirements concerning the
 angular dependence of spin observables, especially at
 forward and back angles, 
   are satisfied and could then be used to deduce
 interesting new dynamics.

\section{The decay distribution}
In photoproduction of a vector meson ($\rho, \phi$) on a nucleon target
the final vector meson decays into two 
pseudoscalar mesons.  The angular distribution of the decay
provides information about the spin-state of the vector meson.
However, only information about the tensor polarizations 
$T^V_{20},T^V_{21},T^V_{22}$ 
can be obtained.
The angular distribution of the pseudoscalar decay  mesons 
 is given by~\cite{KCT}:
\begin{eqnarray}
\overline{W}^V(\bar{\theta},\bar{\phi}) &=& \frac{1}{4 \pi} \xi_V(\bar{\theta})\,
\,
[ 1 -   \sqrt{2}\
T^V_{2\mu}(\Theta,\Phi)\  {\bf C}^*_{2\mu}(\bar{\theta}\bar{\phi})  ] \nonumber
\\
&=& \frac{1}{4 \pi} \xi_V(\bar{\theta})
\, [ 1 -  \sqrt{\frac{1}{2}} T^V_{20} (3 \cos^2 \bar{\theta} - 1) +
\sqrt{3}\ T^V_{2 1}\, \sin 2\bar{\theta} \cos \bar{\phi} -
 \sqrt{3}\, T^V_{2 2}\, \sin^2 \bar{\theta} \cos 2\bar{\phi} ].
\label{WpiPC2}
\end{eqnarray} Here ${\bf C}^*_{2\mu} \equiv
\sqrt{\frac{4 \pi}{5} }\, Y^*_{2\mu} $ is a spherical
harmonic function and the angles $\bar{\theta},\bar{\phi} $ refer to the
 direction between the velocity vector difference, 
 $\Delta \vec{v} \equiv \vec{v}_1-\vec{v}_2, $ and the momentum vector 
of the produced vector meson; $\vec{v}_1$ and $\vec{v}_2$ refer to the
velocity vectors of the two decay mesons in the overall photon-nucleon 
center of mass frame.
 Use of these angles simplifies the  
expression for the angular distribution in the overall photon-nucleon 
center of mass frame in which spin observables are defined.  
Note that the spin observables
$T^V_{2 \mu}$ depend on the vector meson production angles $\Theta, \Phi,$
 as well as on the total c.m. energy.
The factor $\xi,$  which
arises from describing the vector meson decay in the overall 
center of mass system and from a density of state factor, is given by:
\begin{equation}
\xi_V(\bar{\theta}) = \frac{1}{
(\sin^2\bar{\theta} + (\frac{E_\rho}{m_\rho})^2
\cos^2\bar{\theta})^{5/2} }\ \  ,
\label{ksi}
\end{equation} where $m_\rho, E_\rho$ are the vector meson's mass
and energy.

The decay angular distribution 
$\overline{W}^V(\bar{\theta},\bar{\phi})$ does not depend on the
 vector meson's vector polarization 
 and as shown above includes only 
 the vector meson's tensor polarization.  
Once the angular distribution is measured and vector meson rest frame 
density matrices
are provided,  it is necessary to map that data over to the angles 
$\bar{\theta},\bar {\phi}. $ One can then 
project out the vector meson's tensor polarization
 from the normalized ratio 
$\overline{W}^V_{norm}(\bar{\theta},\bar{\phi})
\equiv \overline{W}^V(\bar{\theta},\bar{\phi})/\xi,$   
as described in Sec. V.

The tensor polarization 
must take on values
  that allow the angular distribution 
function $\overline{W}^V(\bar{\theta},\bar{\phi})$
  to be  positive definite.
 By selecting the angles  $\bar{\theta},\bar{\phi}$ one can use that obvious 
 condition to extract allowed limits for the tensor polarization.\footnote{
The allowed ranges for the tensor polarization can also be deduced
by considering the spin-state occupation amplitudes in the pure state
limit.}
In the first and second columns of Table~I a list is given of specific choices
of angles $\bar{\theta},\bar{\phi}$ and the resulting
conditions on $T^V_{20},T^V_{21},T^V_{22}.$ 
Such constraints also arise from direct conditions on the density matrix,
 as will be seen in Sections III and IV.

Thus from Table~I, we see that
 the simple requirement that the decay angular cross section be positive
 yields limits on the
possible tensor polarization.  We now consider other ways
to recognize constraints on the spin observables and the
associated density matrix.  In the next section, we describe the constraints
on $T^V_{20},T^V_{21},T^V_{22}$ that follow from the positivity of the
density matrix.

\section{Limits on observables for a positive definite density matrix}

Recall that for a general observable $\Omega$ the classical ensemble
average is 
\begin{equation}
<\Omega> 
 = \frac{\sum_\alpha \omega_\alpha <\alpha\, | \Omega | \,\alpha>}
{\sum_\alpha \omega_\alpha},
 \end{equation} 
where $\omega_{\alpha}$ is the positive definite probability
for finding a beam particle pointing in the direction stipulated by the
Euler angle label $\alpha.$  
Note that the above is a classical average,
with the quantum effects isolated into the expectation value
for each beam particle $< \,\alpha | \Omega | \,\alpha>.$

The spin density matrix of the vector meson is defined as 
\begin{equation}
\rho = \sum_{\alpha}\ | \,\alpha \,> \omega_{\alpha} < \,\alpha \, |\  ,
\label{density} 
\end{equation} 
where $\omega_{\alpha}$ is non-negative. 
The helicity matrix elements of the density matrix are 
\begin{equation}
\rho_{\lambda \lambda '} = \sum_{\alpha}\  <\lambda\,|\,\alpha> 
\omega_{\alpha} <\alpha\,|\,\lambda'> .
\end{equation} 
The classical ensemble average for observable $\Omega$ is 
now obtained from the density matrix $\rho$ as 
\begin{equation}
<\Omega> = \frac{Tr[\rho \Omega]}{Tr [ \rho ]}. 
\end{equation}

The density matrix $\rho$ is positive definite, which can be 
shown as follows. 
Let us define~\cite{Daboul} a set of vectors $v_{\lambda}$ by its elements
in $\alpha$ space 
\begin{equation}
v_{\lambda}^{\alpha} \equiv <\alpha|\lambda> \sqrt{\omega_{\alpha}}\  . 
\end{equation} 
The elements $\rho_{\lambda \lambda'}$ of the density matrix 
can now be written as
dot-products in $\alpha$ space of the set of vectors $v_{\lambda}$\, , 
\begin{equation}
\rho_{\lambda \lambda'} = (v_{\lambda} , v_{\lambda'}). 
\label{dotproduct}
\end{equation}
Now for any vector $X$ 
\begin{equation}
X^{\dagger} \rho X = 
\sum_{\lambda, \lambda'}\ 
X^*_{\lambda} \rho_{\lambda \lambda'} X_{\lambda'} = 
(\sum_{\lambda } X_{\lambda } v_{\lambda} ,
 \sum_{\lambda'} X_{\lambda'} v_{\lambda'}) \geq 0, 
\label{semiproof}
\end{equation} which displays the positive definiteness of $\rho$. 

At this point we explore the linear constraints on
the matrix elements of $\rho$ implied by Eq.~(\ref{semiproof}).
The density matrix $\rho$ is a $3\times3$ matrix
with elements $\rho_{\lambda,\lambda'}$ where the
helicity $\lambda$ takes the values $ 1,0,-1.$
 Since the production of the vector meson occurs
via a parity conserving mechanism,
the spin density matrix elements satisfy the symmetries
\begin{equation}
\rho_{\lambda \lambda '} = \rho^*_{\lambda ' \lambda},
\ \ \ \ \ {\rm and }\ \ \ \ \
\rho_{\lambda \lambda '} = (-1)^{\lambda - \lambda '}
\rho_{-\lambda -\lambda '} .
\label{Hermiticity}
\end{equation}
The density matrix takes the form
\begin{eqnarray}
\rho  =
\left(\begin{array}{ccc}
 (1 - \rho_{00})/2
& \Re \rho_{10} + i \Im \rho_{10}
& \rho_{1 -1}   \\
  \Re \rho_{10} - i \Im \rho_{10}
& \rho_{00}
&-\Re \rho_{10} + i \Im \rho_{10} \\
  \rho_{1 -1}
&-\Re \rho_{10} - i \Im \rho_{10}
& (1-\rho_{00})/2
       \end{array} \right)  .
\label{rhorhopc-elts}
\end{eqnarray}

In the language of spin observables $P^V_y, T^V_{20}, T^V_{21}, T^V_{22}$
the density matrix can be written as
\begin{equation}
\rho = \frac{1}{3}[ I + \frac{3}{2} \vec{S} \cdot \vec{P}^V
+   \tau \cdot T^V] ,
\label{rhopich}
\end{equation} where $\vec{S}$ is the spin-1
 operator and $\tau$ is the symmetric
traceless rank-2 operator with cartesian components
$\tau_{ij} = \frac{3}{2}(S_i S_j + S_j S_i) - 2 \delta_{ij}$.
In matrix form this becomes
\begin{eqnarray}
\rho  = \frac{1}{3}
\left(\begin{array}{ccc}
1 + \sqrt{\frac{1}{2}} T^V_{20}&
\frac{3}{2}\sqrt{\frac{1}{2}}(-i P^V_y) -
 \sqrt{\frac{3}{2}} T^V_{21}&
\sqrt{3}\ T^V_{22}   \\
\frac{3}{2}\sqrt{\frac{1}{2}}(i P^V_y)-
 \sqrt{\frac{3}{2}} T^V_{21}&
1- \sqrt{2}\ T^V_{20}   &
\frac{3}{2}\sqrt{\frac{1}{2}}(-i P^V_y)
+\sqrt{\frac{3}{2}} T^V_{21} \\
\sqrt{3}\ T^V_{22}&
\frac{3}{2}\sqrt{\frac{1}{2}}(i P^V_y)+
\sqrt{\frac{3}{2}} T^V_{21}  &
1+\sqrt{\frac{1}{2}} T^V_{20}
       \end{array} \right)  .
\label{rhopc-elts}
\end{eqnarray}

If the vector $X$ in Eq.~(\ref{semiproof}) is such that the 
combination $X^*_{\lambda} X_{\lambda'}$ is
symmetric under the exchange of $\lambda $ and $\lambda', $ the obtained constraints 
are similar to the constraints derived in the previous section for 
a positive decay angular distribution. 
The constraints in that case involve 
only the tensor polarizations 
$T^V_{20}, T^V_{21},$ and $T^V_{22}$, and not the vector polarization $P^V_y$. 
This is because symmetric combinations $X^*_{\lambda} X_{\lambda'}$  only pick out the 
symmetric part of the density matrix $\rho_{\lambda \lambda'}$,
 i.~e., in this case the part that
has even rank. The antisymmetric rank-one part of $\rho$, 
which is due to the vector polarization, then gives no contribution 
to $X^{\dagger} \rho X$. This is exactly the symmetry selection made 
if one considers the decay angular distribution of Section II, and 
which was discussed in detail in Ref.~\cite{KCT}. 
The resultant linear constraints are listed in columns 1 and 3 of Table~I. 

Even relations involving the vector polarization $P^V_y$  can be obtained from 
Eq.~(\ref{semiproof}), if the above symmetry restrictions are not
 invoked on $X^*_{\lambda} X_{\lambda'}$. 
However, the resulting 
linear constraints involving $P^V_y$ are
only of academic interest because $P^V_y$ cannot be measured from
the decay angular distribution. 
These additional constraints are therefore not listed in this paper.

\section{Spin observable limits from Schwarz inequalities}

Additional constraints on the density matrix are obtained
using Schwarz inequalities as described in Ref.~\cite{Daboul}.
 Namely, from the previously derived Eq.~(\ref{dotproduct}) and 
\begin{equation}
|(v_{\lambda} , v_{\lambda'})| \le |v_{\lambda}| |v_{\lambda'}|\ ,
\end{equation} 
follows 
\begin{equation}
|\rho_{\lambda \lambda'}| \le 
\sqrt{\rho_{\lambda \lambda}\  \rho_{\lambda' \lambda'}}\ \ . 
\label{Schwarz} 
\end{equation} 
Similar constraints exist for differences or sums of matrix elements 
$\rho_{\lambda \lambda'}$. Such constraints were exploited in 
Ref.~\cite{Daboul}. 
In this case, two additional inequalities can be derived 
 
\begin{equation}
|\rho_{\lambda \lambda'} + \rho_{-\lambda \lambda'}| \le 
\sqrt{2 (\rho_{\lambda \lambda} + \rho_{\lambda -\lambda})\  
\rho_{\lambda' \lambda'}} ,  
\label{PLUS} 
\end{equation} 
and 
\begin{equation}
|\rho_{\lambda \lambda'} - \rho_{-\lambda \lambda'}| \le 
\sqrt{2 (\rho_{\lambda \lambda} - \rho_{\lambda -\lambda})\  
\rho_{\lambda' \lambda'}} .  
\label{MINUS}
\end{equation}

From Eqs.~(\ref{Schwarz},\ref{PLUS},\ref{MINUS}) 
one finds several quadratic constraints.
Using Eq.~(\ref{Hermiticity}) and the property that  
the diagonal matrix elements $\rho_{1,1}, \rho_{0,0}$, and $\rho_{-1,-1}$ 
are non-negative or 
\begin{equation}
1 + \sqrt{\frac{1}{2}}\ T^V_{20} \ge 0 ,
\label{T20bigger} 
\end{equation} 
\begin{equation}
1 - \sqrt{2}\ T^V_{20} \ge 0 ,
\label{T20smaller} 
\end{equation} 
some of the Schwarz inequalities collapse to linear conditions 
\begin{equation}
1 + \sqrt{\frac{1}{2}} T^V_{20} + \sqrt{3} T^V_{22} \ge 0 ,
\label{T22bigger} 
\end{equation} 
\begin{equation}
1 + \sqrt{\frac{1}{2}} T^V_{20} - \sqrt{3} T^V_{22} \ge 0 .
\label{T22smaller} 
\end{equation} 
However, one also finds two very useful 
quadratic conditions that involve the squares 
of $P^V_y$ and $T^V_{21},$  
\begin{equation}
9 (P^V_y)^2 + 12 (T^V_{21})^2 \le  8 (1 - \sqrt{2} T^V_{20}) 
(1 + \sqrt{\frac{1}{2}} T^V_{20}) ,
\label{PyT21}
\end{equation} 
\begin{equation}
9 (P^V_y)^2 + 12 (T^V_{21})^2 \le  4 (1 - \sqrt{2} T^V_{20}) 
(1 + \sqrt{\frac{1}{2}} T^V_{20} -\sqrt{3} T^V_{22}) .
\label{PyT21T22}
\end{equation} The resulting restrictions on the
observables are listed in columns 1 and 4 of Table~I and Table~II.
As one can see, the obtained linear rules are equivalent to those discussed
 earlier.
Again we omit from Table~I linear conditions that include $P^V_y$.

\section{Data Analysis Method}

The Aachen et al.~\cite{Aach} 
collaboration measured the reaction 
$\gamma + p \rightarrow \rho^0 + p$ 
using an 
unpolarized photon beam and an 
unpolarized proton target.
The final proton recoils when 
a $\rho$ meson is produced and the $\rho$
 subsequently decays into 
$\pi^+$ and $\pi^-$ mesons, both of which are detected. 
Hence, the angular pion distribution in the final state is measured.
A fit to this angular distribution 
in the $\rho$ meson rest-frame yields three density matrix elements 
from which we can reconstruct their pion angular distribution in the 
$\rho$ meson rest frame.  
This reconstructed decay pion angular distribution
is called $W^V(\theta ,\phi )$, where $\theta, \phi$ 
define the direction of $\pi^+$ in the $\rho$ rest frame.

The above pion angular distribution depends on the spin state of the 
produced $\rho$ meson, which is described by a 
$3 \times 3$ spin density matrix 
$\rho_{\lambda,\lambda'}$. 
The angular distribution only depends on 
the three real elements $\rho_{00}$, $\Re \rho_{10}$, $\rho_{1-1}$. 
 Values of these elements in the rest frame of the $\rho$ meson 
are published by Aachen et al.~\cite{Aach}
for a set of several photon beam energies and vector meson production angles. 
In order to study reaction mechanisms, however, we are interested in 
spin correlations(single and double spin observables)
 that are defined in the overall center of mass 
frame. How does one obtain spin correlations from these previously 
published density matrix elements?

Our aim is to use the Aachen et al. data in the form of $W^V(\theta ,\phi )$  
to obtain the pion angular distribution in the photon-nucleon center of 
mass frame and re-analyze it in terms of the spin correlations. 
If one or more particles in the reaction are polarized~\cite{KCT},
such future data can be analyzed in a similar way 
 to extract meaningful spin correlations. 
Our procedure is therefore preparation for analysis
of future experimental results 
from Thomas Jefferson Laboratory with polarized photons~\cite{JLAB}. 

The first step is to 
obtain the angular distribution $W^V(\theta ,\phi )$ 
in the $\rho$ meson rest frame from the values of 
the elements $\rho_{00}$, $\Re \rho_{10}$, $\rho_{1-1}$ using 
 \begin{eqnarray}
W^V(\theta,\phi)&=&\frac{3}{4 \pi} 
[
\frac{1 - \, \rho_{00}}{2} + \frac{3 \rho_{00} -1}{2} \cos^2 \theta            
- \sqrt{2}\, \Re  \rho_{10} \sin 2 \theta \cos \phi  
- \rho_{1-1} \sin^2 \theta \cos 2 \phi  
]\ . 
\label{Wpirho} 
\end{eqnarray} Then one 
constructs the angular distribution $\overline{W}^V(\bar{\theta} ,\bar{\phi} )$ 
in the $\gamma$-nucleon center of mass using  
 \begin{eqnarray}
\overline{W}^V(\bar{\theta},\bar{\phi})&=&
\frac{1}{( \sin^2 \bar{\theta} + 
          (\frac{E_{\rho}}{m_{\rho}})^2 \cos^2 \bar{\theta} )^{3/2}}  
W^V(\theta(\bar{\theta}),\phi)\ .
\label{Wbar} 
\end{eqnarray} 
In the c.m. frame the variables are $\bar{\theta}$ and $\bar{\phi}$, 
where $\bar{\theta}, \bar{\phi}$ are the angles 
between the relative velocity of the two decay pions in the $\gamma$-nucleon 
c.m. frame.  These angles are related
to the angles of the decay meson in the vector meson's rest frame by: 
\begin{equation}
\bar{\theta} = \arctan(\frac{E_{\rho}}{m_{\rho}} \tan \theta),
\end{equation} and  $\bar{\phi} = \phi.$ 

 The next step involves including 
the known kinematic 
factor $\xi(\bar{\theta})$  in
$\overline{W}^V(\bar{\theta} ,\phi )$ see Eq.~(\ref{ksi}). 
Aside from the overall factor of
$\xi(\bar{\theta})$, the remaining angular behavior is expressed as a 
series in spherical harmonics:  
\begin{eqnarray}
\overline{W}^V(\bar{\theta},\bar{\phi})&=&\frac{1}{4 \pi} \xi(\bar{\theta}) 
[
(1 - \sqrt{\frac{1}{2}}\, T^V_{20} (3 \cos^2 \bar{\theta} - 1) 
+ \sqrt{3}\, \  T^V_{21} \sin 2\bar{\theta} \cos \bar{\phi} 
-\sqrt{3}\,\ T^V_{22} \sin^2 \bar{\theta} \cos 2\bar{\phi}  
 ]. 
\label{Wpi} 
\end{eqnarray} 
Next we define
 $\overline{W}^V_{norm}(\bar{\theta} ,\phi )$  
using Eq.~(\ref{Wpi}) and  $\overline{W}^V(\bar{\theta} ,\phi ) \equiv 
\xi(\bar{\theta}) \times \overline{W}^V_{norm}(\bar{\theta} ,\phi ).$  

Finally, we project out the three spin 
observables, i.~e., the three tensor polarizations of the vector meson, 
$T^V_{20}, T^V_{21}, T^V_{22}$, 
from $\overline{W}^V_{norm}(\bar{\theta} ,\phi )$ 
 using spherical harmonics $Y_{lm}(\bar{\theta}, \phi)$'s.
These spin observables can then be studied as function of 
photon energy and the vector meson production angles $\Theta, \Phi.$

Once the spin observables are properly defined
 and have correct production angle dependence,
one can visualize the role of the tensor polarization 
in 3D displays of the decay angular distribution.
Examples are given in Fig.~4 for the case of  $\Theta = 0^\circ, 180^\circ$
and positive $T_{20}$ and in Fig. 5 for the case of
  $ \Theta = 0^\circ, 180^\circ $
and negative $T_{20};$  with in both cases $T_{21}=T_{22}=0.$  In Fig. 6,
 a more realistic case, based on Ref.~\cite{Aach} for $E_\gamma = 3 GeV$
and  $\Theta =70^\circ$ is shown; namely,
for $T_{20}=-.72, T_{21}=-.21, T_{22}=.19.$ One can therefore associate
a shape of $\overline{W}$  with each point in the
allowed $T_{20},T_{21},T_{22}$ space.

In the process of carrying out these steps, 
 we found four cases of the Aachen data that
do not satisfy the constraints in Tables~I and II. Therefore
 those sets had to be rejected.  Subsequently we found
another author had also rejected some of
the data~\cite{Daboul} using similar general constraints.

  It would therefore be best to
incorporate these constraints directly into the data analysis.

\section{CONCLUSION}

For analysis of experimental data for vector meson photoproduction one 
should describe the resulting angular decay distribution in the overall 
$\gamma N$ c.m. frame instead of in the commonly used vector-meson 
rest frame. In the c.m. frame, one should use the angles 
$\bar{\theta}, \bar{\phi}$
of the relative velocity vector 
$\Delta \vec{v} = \vec{v}_1 - \vec{v}_2$.
In the transformation from the vector-meson rest frame to the 
$\gamma N$ c.m. frame, and due to the use of the angles 
$\bar{\theta}, \bar{\phi},$ 
a kinematical factor $\xi$ of Eq.~(\ref{ksi}) needs to be included.
Furthermore, constraints should be satisfied by the observables
$T^V_{20},T^V_{21},T^V_{22}$. 
All of these constraints can be derived from the 
positivity of the density matrix. 

We have also explored restrictions which follow from positivity of the 
eigenvalues of the spin density matrix $\rho$ such as the conditions that
 $\det \rho \geq 0$ 
and Trace$[\rho^2] \leq 1$,  see Ref.~\cite{Minnaert}.
The relations obtained from these restrictions
 are respectively cubic and quadratic in the spin observables and involve
the vector meson's vector polarization $P^V_y$. 
A more complete set of relations can be obtained
by exploring the explicit forms of the roots 
$x_1, x_2, x_3$ of the 
eigenvalue equation of $\rho.$ 
We impose the conditions that $0 \leq x_i \leq 1,$
 $x_1 + x_2 + x_3 =1,$ 
and $x^2_1 + x^2_2+x^2_3 \leq 1.$ 
For example, one root is 
$x_1 = \frac{1}{3} ( 1 + \frac{1}{\sqrt{2}} T^V_{20} + \sqrt{3} T^V_{22})$.
From $0\leq x_1\leq 1$ 
relations follow that are similar to the rules in Table~I. 
The other two roots are
$x_2 = \frac{1}{3} ( 1 - \frac{1}{2 \sqrt{2}} T^V_{20} -
\frac{\sqrt{3}}{2} T^V_{22}) -  \frac{1}{2} \sqrt{P_y^2 +
\frac{4}{3}T_{21}^2 + ( \frac{1}{\sqrt{2}} T_{20} -
\frac{1}{\sqrt{3}} T_{22})^2}$
and
$x_3 = 1 - x_1 - x_2$.
Both $x_2$ and $x_3$ involve $P^V_y$ and the above root rules lead to 
quadratic constraints that are included in Table~II. 

The linear and quadratic constraints on 
$T^V_{20}, T^V_{21}, T^V_{22}$ mean that the allowed domain in 
the 3-dimensional $T^V_{20} - T^V_{21} - T^V_{22}$ space is confined. 
As examples Figs. (1-3) show the allowed areas for the 
2-dimensional subspaces 
$T^V_{20} - T^V_{21}$ , $T^V_{20} - T^V_{22}$ , 
and $T^V_{22} - T^V_{21}$ respectively. 
The dashed lines indicate the various upper bounds and the solid lines 
represent the lower bounds associated with the various constraints.
The allowed regions are shaded. 
Figs. (1-3) refer in each case
 to 2-dimensional subspaces and only those constraints 
are shown that are independent of the values of the observable that would 
play the role of the third dimension. The full 3-dimensional 
representation of constraints is richer.

We close with some remarks about double spin observables. 
For a polarized photon beam again the angular distribution 
of the decay mesons can be measured. This decay distribution now depends 
on the single spin observables 
$T^V_{20}, T^V_{21}, T^V_{22}$ as well as on the double spin correlations 
$C^{\gamma V}_{x20}, C^{\gamma V}_{x21}, C^{\gamma V}_{x22}, 
C^{\gamma V}_{y21}, C^{\gamma V}_{y22}, 
C^{\gamma V}_{z21}, C^{\gamma V}_{z22}$ (Ref.~\cite{KCT}). 
(Again the vector polarization and correlations with the vector polarization 
cannot be measured.) 
Similar to the method described in Sections II and III and based on positive 
decay angular distributions and positivity of the complete density matrix 
for this case, linear and quadratic relations involving double spin observables 
can be derived. 

The constraints on the spin observables presented in this paper
 should be incorporated directly into the analysis 
 of forthcoming data.

\acknowledgments
The authors wish to thank
Mr. Wen-tai Chiang for his help at an early
stage of this study. 
One author (W.M.K.) thanks the University of Pittsburgh,
another (F.T.) thanks Rutgers University for
warm hospitality.  This research was
supported, in part, by the U.S. National Science Foundation
Phy-9504866(Pitt) and Phy-9722088(Rutgers).

%

\newpage
 
\begin{table}

\protect\label{tab:T2mconstraints}
\begin{center}
\begin{tabular}{ c | c | c | c  }
 Constraints  & {$W \geq 0$} & $X^{\dagger} \rho X \geq 0$ &  Schwarz \\ 
\hline
& & &   \\
  $ 1 - \sqrt{2} T_{20} \geq 0$ 
& $ \bar{\theta}=0$ \ \ 
& $ \rho_{00} \geq 0$ \ \ 
& $ \rho_{00} \geq 0$ \\  
& & &   \\
  $ 1 + \frac{1}{\sqrt{2}} T_{20} \geq 0$ 
& $ \bar{\theta}=\frac{\pi}{2};\bar{\phi}=\frac{\pi}{4}$\ \ 
& $ \rho_{11} \geq 0$ \ \ 
& $ \rho_{11} \geq 0$ \\  
& & &   \\
\hline
\hline
& & &   \\
  $ 1 + \frac{2}{\sqrt{3}} T_{21} \geq 0$ 
& $ \bar{\theta} = \bar{\theta}_c; \bar{\phi}= \frac{\pi}{4}$ \ \   
& $ 1 - 2 \sqrt{2} \Re \rho_{10} \geq 0 $ 
& \\  
& & &   \\
  $ 1 - \frac{2}{\sqrt{3}} T_{21} \geq 0$ 
& $ \bar{\theta}=\bar{\theta}_c ; \bar{\phi}= \frac{3 \pi}{4}$ \ \   
& $ 1 + 2 \sqrt{2} \Re \rho_{10} \geq 0 $ 
& \\  
& & &   \\
\hline
\hline
& & &   \\
  $ 1 + \frac{2}{\sqrt{3}} T_{22} \geq 0$ 
& $ \bar{\theta}=\bar{\theta}_c; \bar{\phi}= \frac{\pi}{2}$ \ \   
& $ 1 + 2 \rho_{1 -1} \geq 0 $
& \\  
& & &   \\
\hline
\hline
& & &   \\
  $ 1 + \frac{1}{\sqrt{2}} T_{20} + \sqrt{3} T_{22} \geq 0$ 
& $ \bar{\theta}=\frac{\pi}{2};\bar{\phi}=\frac{\pi}{2}$  
& $ 1 - \rho_{00} + 2 \rho_{1 -1} \geq 0 $
& $ |\rho_{1 -1}|^2 \leq \rho_{11} \rho_{-1 -1}$ \\ 
& & &   \\
  $ 1 + \frac{1}{\sqrt{2}} T_{20} - \sqrt{3} T_{22} \geq 0$ 
& $ \bar{\theta}=\frac{\pi}{2};\bar{\phi}=0$  
& $ 1 - \rho_{00} - 2 \rho_{1 -1} \geq 0 $
& $ |\rho_{1 -1}|^2 \leq \rho_{11}  \rho_{-1 -1}$ \\  
& & &   \\
  $ 2 - \frac{1}{\sqrt{2}} T_{20} - \sqrt{3} T_{22} \geq 0$ 
& 
& 
& see caption(*)\\ 
& & &   \\
  $ 2 - \frac{1}{\sqrt{2}} T_{20} + \sqrt{3} T_{22} \geq 0$ 
& $ \bar{\theta} = \frac{\pi}{4}; \bar{\phi}= \frac{\pi}{2}$  
& 
& \\   
& & &   \\
\hline
\hline
& & &   \\
  $ 1 - \frac{1}{2 \sqrt{2}} T_{20} + 
        \sqrt{\frac{3}{2}} T_{21}\geq 0$ 
& $ \bar{\theta}=\frac{\pi}{4}; \bar{\phi}= \frac{\pi}{4}$ \ \   
& $ 1 + \rho_{00} - 4 \Re \rho_{10} \geq 0 $
& \\  
& & &   \\
  $ 1 - \frac{1}{2 \sqrt{2}} T_{20} - 
        \sqrt{\frac{3}{2}} T_{21}\geq 0$ 
& $ \bar{\theta}=\frac{3 \pi}{4}; \bar{\phi}= \frac{\pi}{4}$ \ \   
& $ 1 + \rho_{00} + 4 \Re \rho_{10} \geq 0 $
& \\  
& & &   \\
\hline
\hline
& & &   \\
  $ 1 + \frac{2 \sqrt{2}}{\sqrt{3}} T_{21} - 
        \frac{2}{\sqrt{3}} T_{22}\geq 0$ 
& $ \bar{\theta}=\bar{\theta}_c; \bar{\phi}= 0$ \ \   
& $ 1 - 4 \Re \rho_{10} - 2 \rho_{1 -1} \geq 0 $
& \\  
& & &   \\
  $ 1 - \frac{2 \sqrt{2}}{\sqrt{3}} T_{21} - 
        \frac{2}{\sqrt{3}} T_{22}\geq 0$ 
& $ \bar{\theta}=\bar{\theta}_c; \bar{\phi}= \pi$ \ \   
& $ 1 + 4 \Re \rho_{10} - 2 \rho_{1 -1} \geq 0 $
&   \\
\end{tabular}
\end{center}
\caption{ Linear Constraints for $T_{20}, T_{21}, T_{22}$; 
$\bar{\theta}_c = \arccos(\frac{1}{\sqrt{3}})$. 
(*) This condition follows from $ x_1 \leq 1.$}
\end{table}

\newpage
 
\begin{table}

\protect\label{tab:PyT2mconstraints}
\begin{center}
\begin{tabular}{ l | c | c }
 Constraints  & Schwarz Inequality & Other \\ 
\hline
& &  \\
\noindent   $ 9 P_y^2 + 12 T_{21}^2 \leq  
    8 (1 - \sqrt{2} T_{20}) ( 1 +\frac{1}{\sqrt{2}} T_{20}) $ 
& $ |\rho_{1 0}|^2 \leq \rho_{11} \rho_{0 0} $
& 
\\[5pt]
& &  \\
\hline
& &  \\
$ 9 P_y^2 + 12 T_{21}^2 \leq   
    4 (1 - \sqrt{2} T_{20}) ( 1 +\frac{1}{\sqrt{2}}
    T_{20} - \sqrt{3} T_{22}) $ 
& $ |\rho_{1 0}- \rho_{-1 0}|^2 \leq  
    2| \rho_{11}-\rho_{1 -1 }| \rho_{00}$
& $ x_2 \geq 0$
\\[5pt]
& &  \\
\hline
& &  \\
  $ 9 P_y^2 + 12 T_{21}^2 \leq   
    8 (1 + \frac{1}{\sqrt{2}} T_{20}) ( 2 -\frac{1}{\sqrt{2}}
    T_{20} + \sqrt{3} T_{22}) $ 
& 
& $ x_3 \leq 1$
\\[5pt]
& &  \\
\hline
& &  \\
  $ 9 P_y^2 + 12 T_{21}^2 + 12 T_{22}^2 + 6 T_{20}^2  \leq  12 $ 
& 
& $Tr[\, \rho^2\, ] \leq 1 $ \\
& &    \\
\end{tabular}
\end{center}
\caption{ Quadratic constraints for $P_y, T_{20}, T_{21}, T_{22}$ }
\end{table}

\newpage
\begin{figure}[t]
\epsfxsize= 5in 
\epsfbox{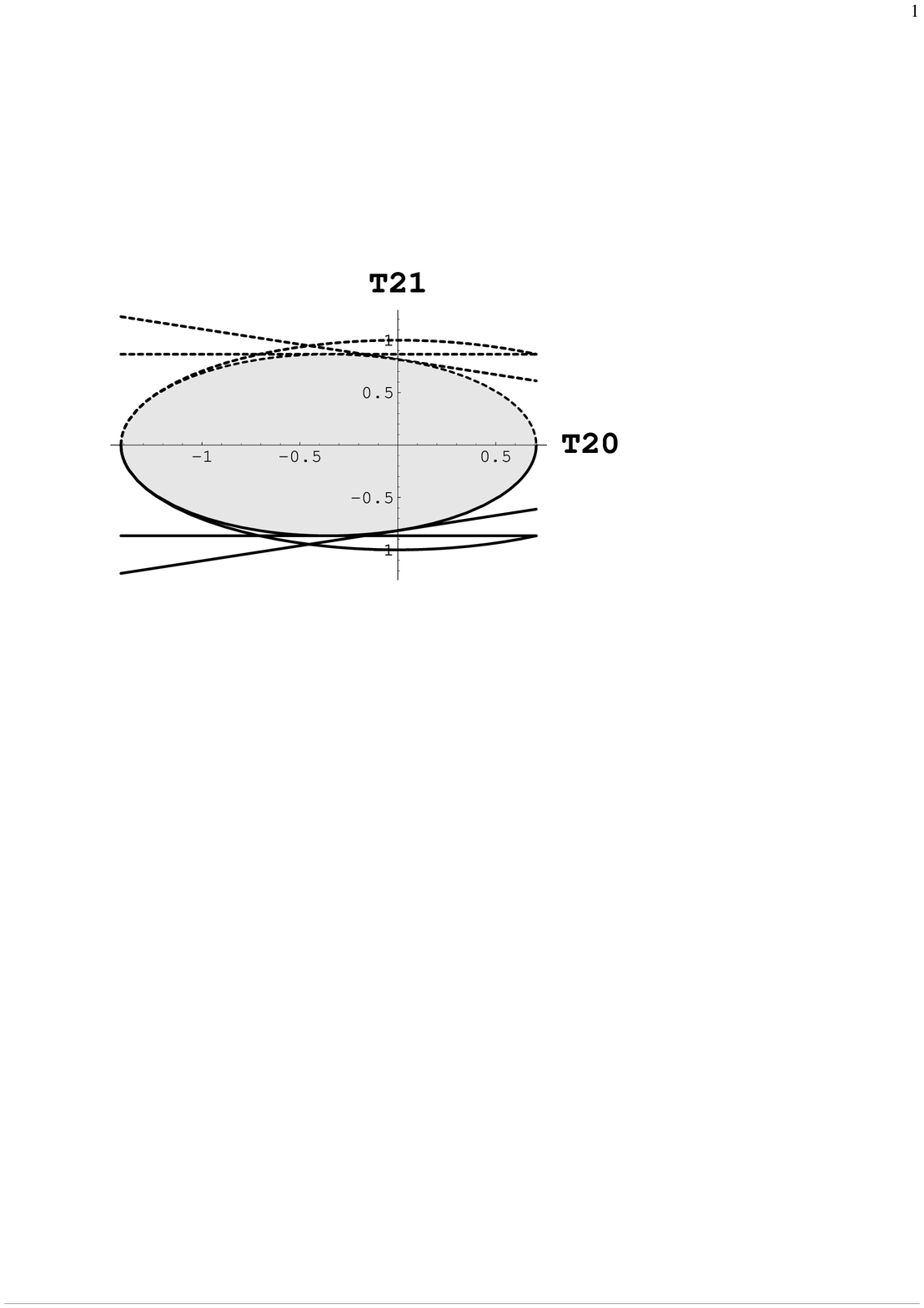}    
\vskip -3in
\caption{ 
Allowed domain in the space spanned by $T^V_{20}$ and $T^V_{21}$ 
independent of the value of $T^V_{22}$. 
Dashed lines are upper bounds. Solid lines are lower bounds. 
Linear constraints are from Table 1 and quadratic constraints
are from Table II.  The shaded area represents the allowed region.
} 
\end{figure}

\vskip -2.0in
\begin{figure}[t]
\epsfxsize= 5in 
\epsfbox{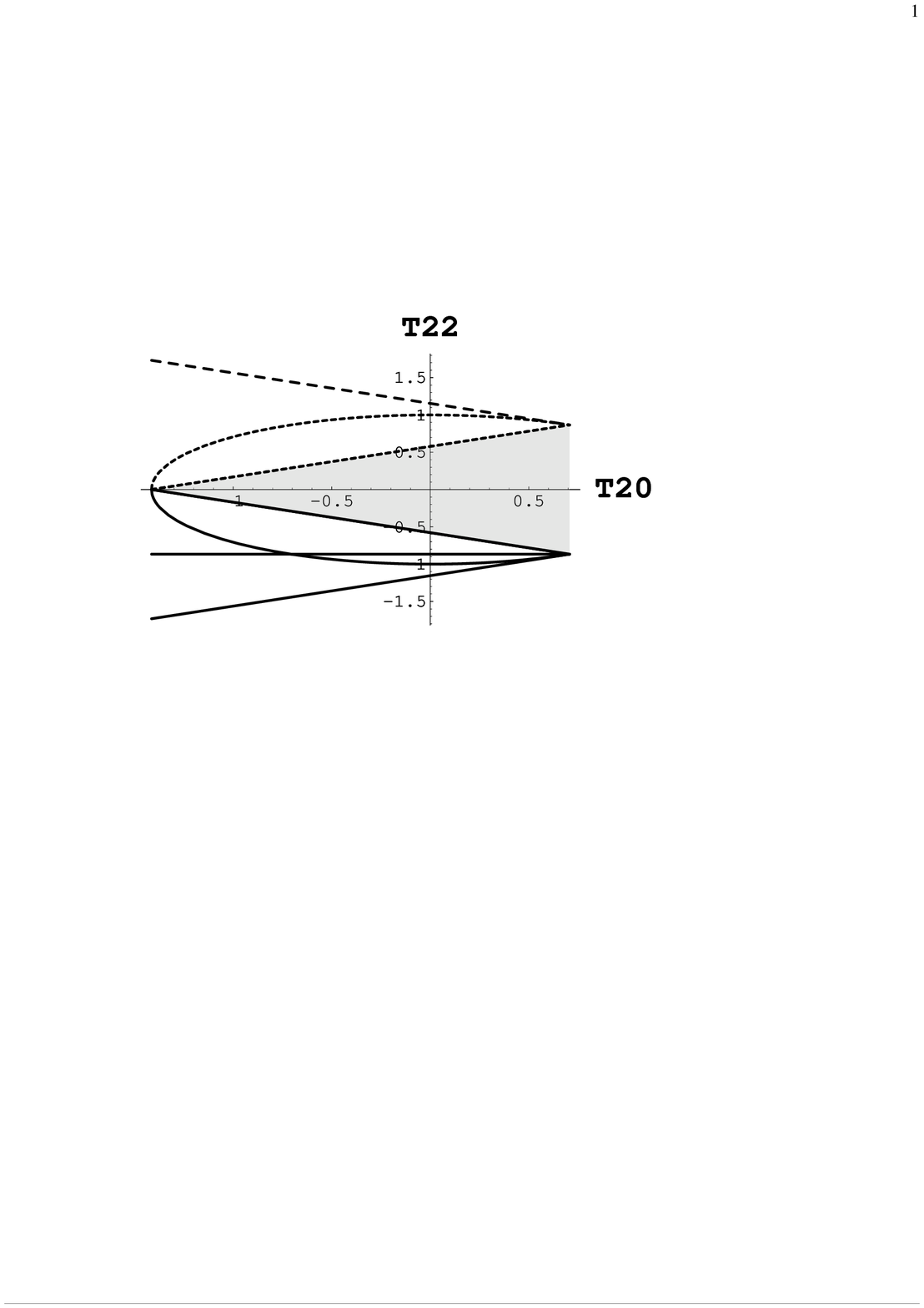}    
\vskip -3in
\caption{ 
Allowed domain in the space spanned by $T^V_{20}$ and $T^V_{22}$ 
independent of the value of $T^V_{21}$. 
Dashed lines are upper bounds. Solid lines are lower bounds. 
Linear constraints are from Table 1 and quadratic constraints
are from Table II.  The shaded area represents the allowed region.
} 
\end{figure}

\vskip -2.0in
\begin{figure}[t]
\epsfxsize= 5in 
\epsfbox{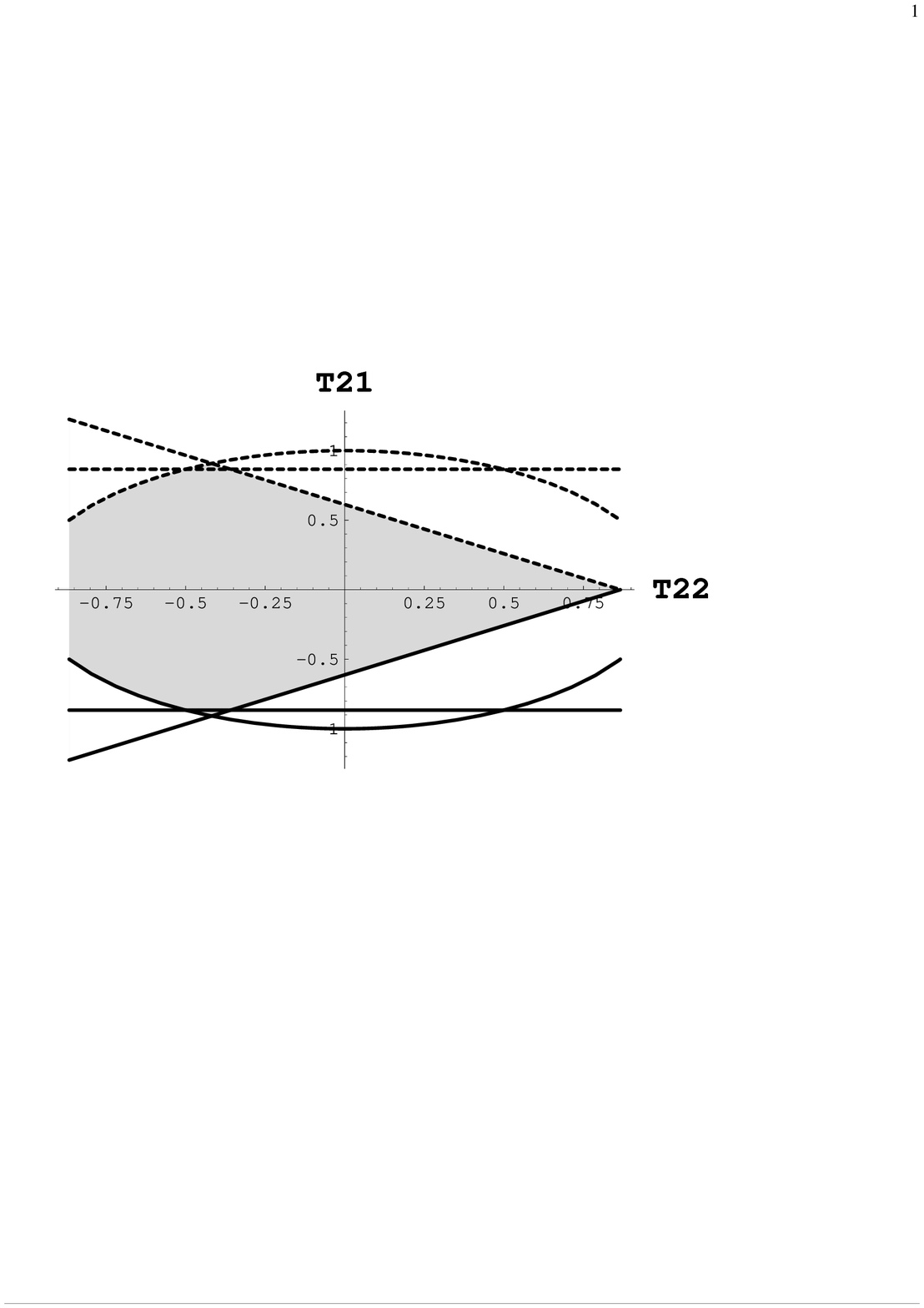}    
\vskip -2.5in
\caption{ 
Allowed domain in the space spanned by $T^V_{22}$ and $T^V_{21}$ 
independent of the value of $T^V_{20}$. 
Dashed lines are upper bounds. Solid lines are lower bounds. 
Linear constraints are from Table 1 and quadratic constraints
are from Table II.  The shaded area represents the allowed region.
} 
\end{figure}

\newpage
\begin{figure}[t]
\epsfxsize= 5in 
\epsfbox{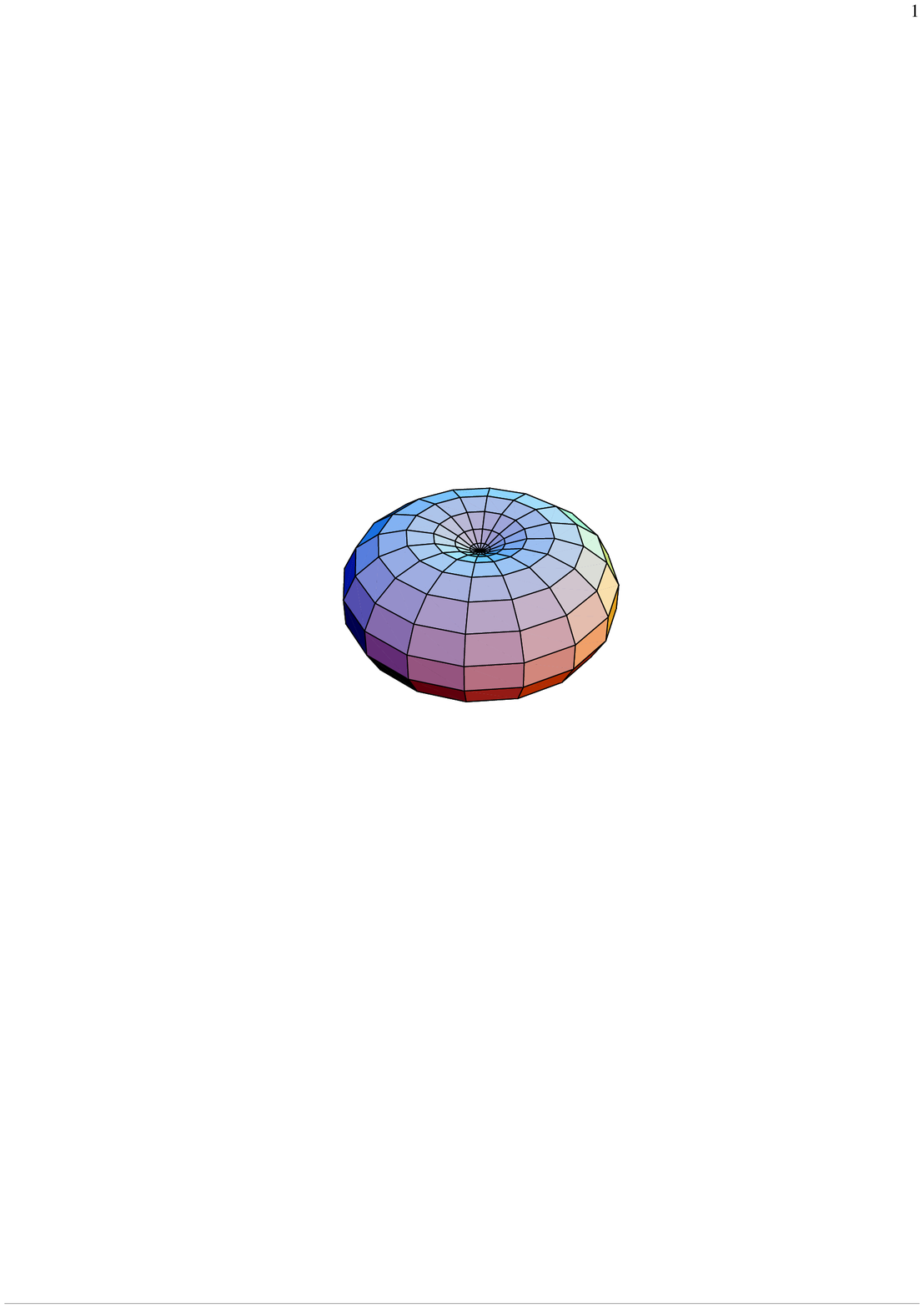}    
\vskip -3in
\caption{  Forward/Back $\overline{W}$ for $T_{20}>0.$
This is the distribution in terms of the
angles $\bar{\theta}, \bar{\phi}$ with the up direction
being the direction of the produced vector meson's momentum.}
\end{figure}
\vskip -2.0in
\begin{figure}[t]
\epsfxsize=5in 
\epsfbox{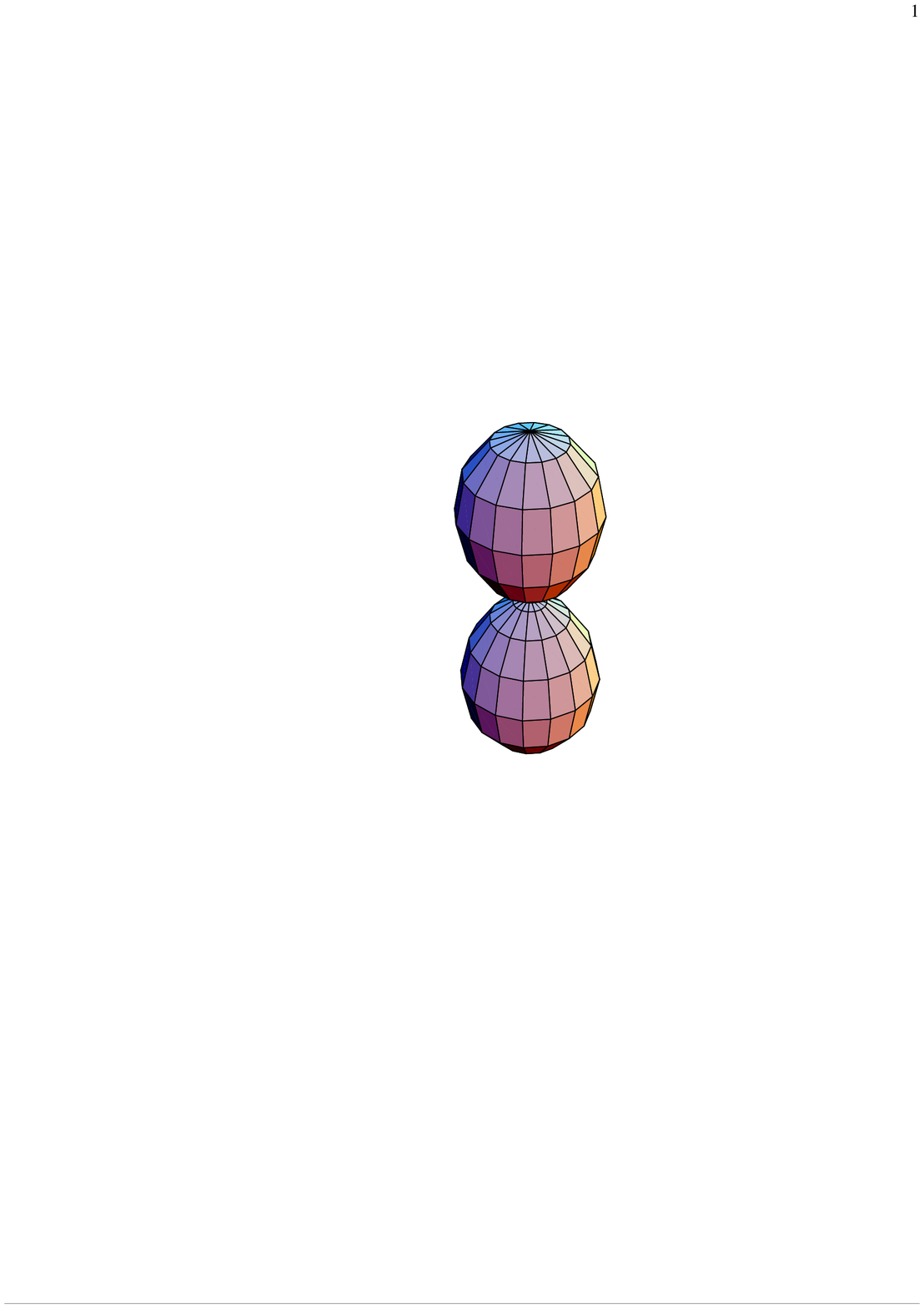}    
\vskip -2.5in
\caption{ Forward/Back $\overline{W}$ for $T_{20}<0.$
This is the distribution in terms of the
angles $\bar{\theta}, \bar{\phi}$ with the up direction
being the direction of the produced vector meson's momentum.}
\end{figure}
\vskip -2.0in
\begin{figure}[t]
\epsfxsize= 5in 
\epsfbox{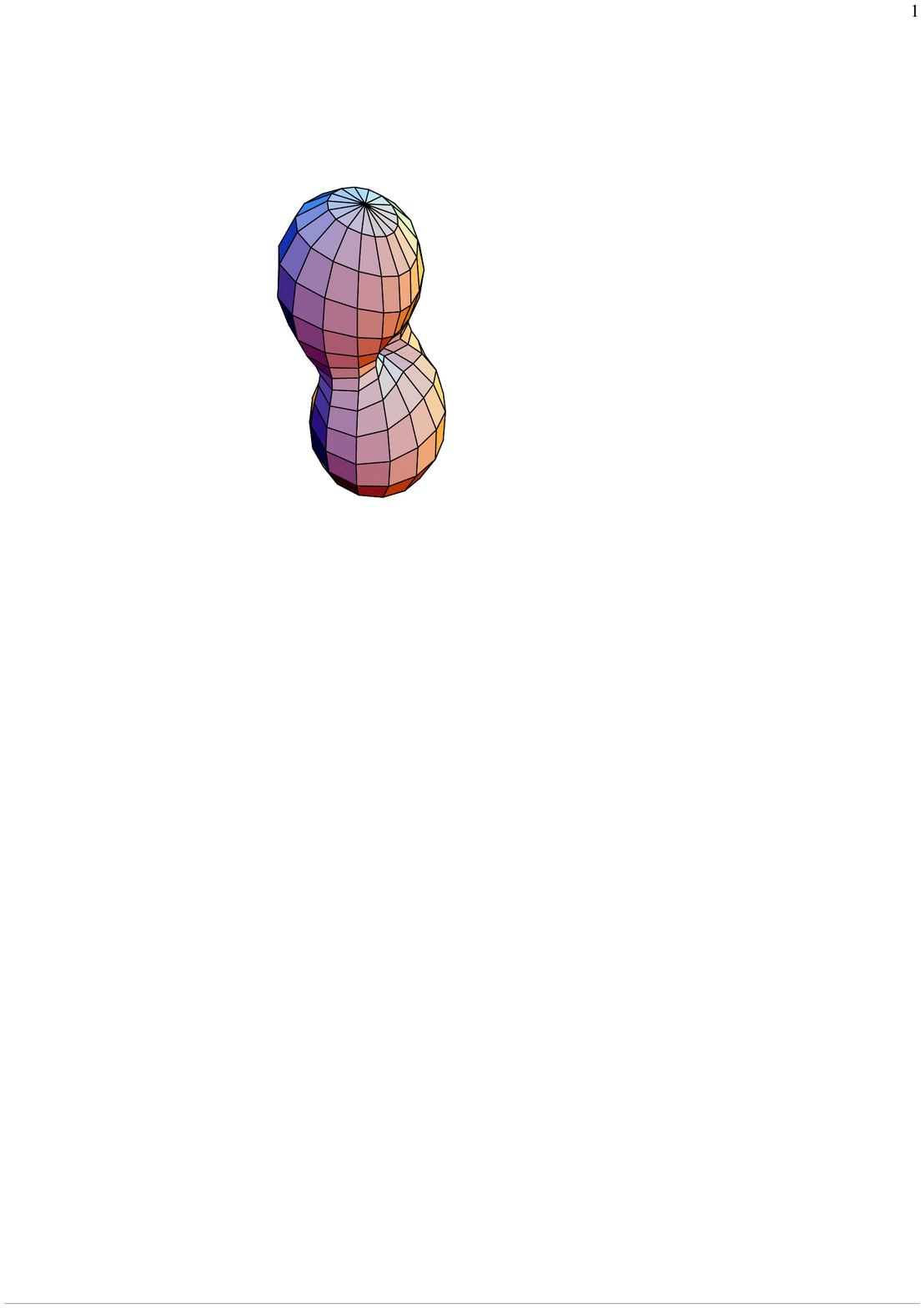}    
\vskip -3in
\caption{ Shape of $\overline{W}$ for $ T_{20}=-.72, T_{21}=-.21, T_{22}=.19.$
This is the distribution in terms of the
angles $\bar{\theta}, \bar{\phi}$ with the up direction
being the direction of the produced vector meson's momentum. }
\end{figure}


\begin{references}
%
\bibitem{KCT} W.~M.~ Kloet. Wen-tai Chiang and F. Tabakin, 
Phys. Rev. {\bf C 58}, 1086 (1998).
%
%
\bibitem{PST} M.~Pichowsky, C.~Savkli, and F.~Tabakin, 
Phys. Rev. {\bf C 53 }, 593  (1996).
%
%
\bibitem{Aach} Aachen-Berlin-Bonn-Hamburg-Heidelberg-Munchen Collaboration, 
Phys. Rev. {\bf 175}, 1669 (1968).
%
\bibitem{Daboul} J. Daboul, 
Nucl. Phys. B {\bf 4}, 180 (1967).
%
\bibitem{Minnaert} P. Minnaert, 
Phys. Rev. {\bf 151}, 1306 (1966).
%
\bibitem{CT} Wen-tai Chiang and F.~Tabakin, 
Phys. Rev. {\bf C  },   (199).
%
\bibitem{JLAB}   
E-94-109 P. Cole, R. Whitney, and J. Connelly,
Photoproduction of the Rho Meson from the Proton with Linearly
Polarized Photons;
E-98-109 D. Tedeschi, P. Cole, and J. Mueller,
Photoproduction of phi Mesons with Linearly Polarized Photons;
E-93-031 C. Marchand, M. Anghinolfi, and J. Laget,
 Photoproduction of Vector Mesons at High t;
E-93-033 J. Napolitano and D. Weygand,  A Search for Missing
Baryons Formed in $\gamma P \rightarrow P \pi^+ \pi^- $
 Using the CLAS at CEBAF.
%
\end{references}
\end{document}